\documentclass[aps,prd,preprint,tightenlines,superscriptaddress,
amsfonts,amssymb,amsmath,byrevtex,showpacs,nofootinbib]{revtex4-1}
\usepackage[polish,english]{babel}
\usepackage[applemac]{inputenc}
\usepackage[T1]{fontenc}
\usepackage{latexsym}
\usepackage{graphicx}
\usepackage{geometry}
\usepackage{epstopdf}
\usepackage{color}

\usepackage{xcolor}
\usepackage{pdfsync}
\usepackage{fancyhdr}
\usepackage{makeidx}
\usepackage[breaklinks,colorlinks,backref]{hyperref}
\hypersetup{
    colorlinks, 
    linktoc=all, 
    linkcolor=red,  
  citecolor=hyptxt,
  urlcolor=blue}
  \hypersetup{
  citebordercolor=,
  filebordercolor=green,
  linkbordercolor=blue
}
\definecolor{hyptxt}{rgb}{0.7, 0.4, 0.9}
\usepackage{bibtopic}

\newcommand{\dR}{\mathbb R}
\newcommand{\dC}{\mathbb C}

\newcommand{\id}{\mathbb I}

\newcommand{\be}{\begin{equation}}
\newcommand{\ee}{\end{equation}}

\newcommand{\I}{\mathbb I}

\newcommand{\R}{\mathbb R}

\newcommand{\ket}[1]{|\kern.3ex#1\kern.3ex\rangle}
\newcommand{\bra}[1]{\langle\kern.3ex #1 \kern.3ex|}
\newcommand{\scalar}[2]{\langle\kern.3ex #1 \kern.3ex|\kern.3ex#2\kern.3ex\rangle}
\newcommand{\norm}[1]{\|\kern.3ex#1\kern.3ex \|}

\newcommand{\UnitOp}{\hat{\mathbb{I}}} 
\newcommand{\Group}[1]{\mathrm{#1}} 
\newcommand{\Bra}[1]{\langle #1 \vert} 
\newcommand{\Ket}[1]{\vert #1 \rangle} 
\newcommand{\BraKet}[2]{\langle #1 \vert #2 \rangle} 
\newcommand{\Aver}[1]{\langle #1 \rangle}

\begin{document}

\title{Dynamical structures specific to strong gravitational field: \\
  Quantum formalism}


\author{Andrzej G\'{o}\'{z}d\'{z}}
\email{andrzej.gozdz@umcs.lublin.pl}
\affiliation{Institute of Physics, Maria Curie-Sk{\l}odowska
University, pl.  Marii Curie-Sk{\l}odowskiej 1, 20-031 Lublin, Poland}

\author{W{\l}odzimierz Piechocki} \email{wlodzimierz.piechocki@ncbj.gov.pl}
\affiliation{Department of Fundamental Research, National Centre for Nuclear
  Research, Ho{\.z}a 69, 00-681 Warszawa, Poland}

\author{Grzegorz Plewa} \email{grzegorz.plewa@ncbj.gov.pl}
\affiliation{Department of Fundamental Research, National Centre for Nuclear
  Research, Ho{\.z}a 69, 00-681 Warszawa, Poland}

\date{\today}

\begin{abstract}

  The dynamics of the general Bianchi IX spacetime, near the gravitational
  singularity, underlies the Belinskii, Khalatnikov and Lifshitz scenario.
  Asymptotically, near the singularity, the oscillations of the directional
  scale factors (defining the spacetime metric) are freezed so the evolution of
  the Bianchi IX model is devoid of chaotic features.  However, it includes
  special structures, that we call the wiggles, which change their properties
  in the asymptotic regime. We propose the formalism, based on an affine
  quantization scheme, to examine the fate of the wiggles at the quantum
  level. We present the way of comparing the classical wiggles with their
  quantum counterparts. We expect that our work may contribute towards
  understanding some quantum aspects of the BKL scenario.

\end{abstract}

\pacs{04.60.-m, 04.60.Kz, 0420.Cv}

\maketitle

\tableofcontents

\section{Introduction}

The Belinskii, Khalatnikov and Lifshitz (BKL) scenario is thought to describe a
generic solution to the Einstein equations near either spacelike
\cite{BKL22,BKL33} or timelike \cite{P1,P2} gravitational singularities. This
scenario was obtained by the generalization of the dynamics of the general
Bianchi IX model.

It is well known that the dynamics of the vacuum Bianchi IX includes the
oscillations of the dynamical scale factors \cite{BKL22}, which underly the
chaotic dynamics of that evolution (see, e.g., \cite{Cornish:1996yg}). If one
makes the analysis of the Bianchi IX dynamics corresponding to the general
case, when the 3-metric of space cannot be diagonalized globally once for all
moments of time\footnote{In the vacuum case the 3-metric can be globally
  diagonalized.}, one can find that asymptotically near the singularity the
oscillations are freezed so the evolution is devoid of the chaotic behaviour
\cite{ryan,bel,Czuchry:2014hxa}. However, there are other interesting
structures, which we call the wiggles, specific to this dynamics.

The existence of physical structures in the evolution of spacetime is of
fundamental importance since if they exist, they can be treated as seeds of
real structures in the Universe like, e.g., inhomogeneities of the CMB
spectrum. The latter is believed to be the origin of large scale structures,
like galaxies and clusters of galaxies, visible presently on the sky.  On the
other hand, if quantization does not erase these structures, they can be used
to get insight into quantum fluctuations underlying quantum gravity expected to
play an important role near the cosmological and astrophysical singularities.
This paper is devoted to the construction of the formalism to be used in the
examination of the dynamical gravitational structures at the quantum level.

Our paper is organized as follows: In Sec. \!II we recall (to make our paper
self-consistent) the Hamiltonian formulation of our gravitational system.
Section III is devoted to the construction of the quantum formalism.  It is
based on using the affine coherent states ascribed to the physical phase space
and the resolution of the unity in the carrier space of the unitary
representation of the affine group. The quantum dynamics is defined in
Sec. \!IV.  Section V concerns the probability measure defined on the physical
phase space.  An outline of how the formalism might be applied is presented in
Sec. \!VI.

\section{Classical dynamics}

The asymptotic form (near the singularity) of the dynamical equations of the
general (nondiagonal) Bianchi IX model is the following \cite{ryan,bel}
\begin{equation}\label{a1}
 \frac{d^2\ln a}{d T^2}  = b/a - a^2,~~~\frac{d^2\ln b}{d T^2}
= a^2-b/a + c/b,~~~\frac{d^2\ln c}{d T^2} =
 a^2 - c/b,
\end{equation}
where $a,b,c$ are functions of an evolution parameter $T$, and are interpreted
as the directional scale factors of considered anisotropic universe.  The
solution to \eqref{a1} should satisfy the dynamical constraint
\begin{equation}\label{a2}
 \frac{d\ln a}{d T}\;\frac{d\ln b}{d T}
+ \frac{d\ln a}{d T}\;\frac{d\ln c}{d T}
+ \frac{d\ln b}{d T}\; \frac{d\ln c}{d T} = a^2 + b/a + c/b.
\end{equation}
Eqs. \!\eqref{a1} and \eqref{a2} define a nonlinear coupled system of
differential equations.  We recommend Sec. 6 of Ref.  \cite{bel} for the
derivation of these equations from the exact dynamics.

Recently, we have derived the reduced phase space formulation of the classical
dynamics corresponding to the above dynamics  \cite{Czuchry:2012ad}.  The two form $\Omega$
defining the Hamiltonian formulation is given by
\begin{equation}\label{sim1}
\Omega = dq_1 \wedge dp_1 + dq_2 \wedge dp_2 + dt \wedge dH,
\end{equation}
where the variables $(q_1, q_2,p_1, p_2)$ parameterise the physical phase
space, $H= H(q_1, q_2,p_1, p_2, t)$ is the Hamiltonian generating the classical
dynamics, and $t$ is an evolution parameter specific to the Hamiltonian
formulation. There is no explicit relation between the evolution parameters $T$
and $t$. The variables $q_1$ and $q_2$ are related to $a$ and $b$ via $q_1 =
\ln{a}$ and $q_2 = \ln{b}$, and the variable $c$ is expressed in terms of other
variables to implement the constraint \eqref{a2} into the Hamiltonian scheme.

The Hamiltonian reads \cite{Czuchry:2012ad}
\begin{equation}\label{sim3}
H(q_1, q_2, p_1, p_2; t)
= q_2+ \ln \left[-e^{2 q_1}-e^{q_2 - q_1}-\frac{1}{4}(p_1^2
+ p_2^2 +t^2) +\frac{1}{2}(p_1 p_2 + p_1 t + p_2 t)\right],
\end{equation}
and Hamilton's equations are:
\begin{eqnarray}
\label{ff1}
\frac{d q_1}{d t} &=& \frac{\partial H}{\partial p_1}
=  \frac{ p_2 - p_1 + t }{2 F},\\
\label{ff2}
\frac{d q_2}{d t} &=& \frac{\partial H}{\partial p_2}
=  \frac{ p_1 - p_2 + t }{2 F},\\
\label{ff3}
\frac{d p_1}{d t} &=& - \frac{\partial H}{\partial q_1}
= \frac{2 e^{2 q_1}- e^{ q_2 - q_1 }}{F}  ,\\
\label{ff4}
\frac{d p_2}{d t} &=&- \frac{\partial H}{\partial q_2}
= - 1 + \frac{e^{q_2 - q_1}}{F},
\end{eqnarray}
where
\begin{equation}\label{eq1}
F(q_1,q_2,p_1,p_2, t):= -e^{2 q_1}-e^{q_2 -q_1}
-\frac{1}{4}(p_1^2 + p_2^2 +t^2)+\frac{1}{2}(p_1 p_2 + p_1 t + p_2 t)> 0.
\end{equation}
The constraint in the r.h.s. of \eqref{eq1} is not of dynamical origin; it
results from the restriction of the dynamics to the physical phase space
parameterized by real (not complex) variables.  Equations
\eqref{ff1}--\eqref{ff4} define a coupled system of nonlinear ordinary
differential equations. The solution defines the physical phase space of our
gravitational system. The regularized version of the Hamiltonian \eqref{sim3}
(to be used in calculations) is presented in App. \! \ref{regular}.

Applying the simple algebraic identity $(A + B + C)^2 = A^2 + B^2 + C^2 + 2AB +
2BC + 2AC$ to Eq. \eqref{eq1} gives:
\begin{align}
F(q_{1}, q_{2}, p_{1}, p_{2}, t) &= -e^{2q_{1}} - e^{q_{2} - q_{1}}
- \frac{1}{4}(p_1 - p_2 + t)^2 + p_1 t , \label{eq2} \\
F(q_{1}, q_{2}, p_{1}, p_{2}, t) &= -e^{2q_{1}} - e^{q_{2} - q_{1}}
- \frac{1}{4}(-p_1 + p_2 + t)^2 + p_2 t \label{eq3}, \\
F(q_{1}, q_{2}, p_{1}, p_{2}, t) &= -e^{2q_{1}} - e^{q_{2} - q_{1}}
- \frac{1}{4}(p_1 + p_2 - t)^2 + p_1 p_2 \label{eq4}.
\end{align}
Combining \eqref{eq1} and \eqref{eq2} we get
\begin{equation}\label{eq5}
p_1 > \frac{1}{t}\left[e^{2q_{1}} + e^{q_{2} - q_{1}}
+ \frac{1}{4}(p_1 - p_2 + t)^2\right] ,
\end{equation}
whereas \eqref{eq1} and \eqref{eq3} give
\begin{equation}\label{eq6}
p_2 > \frac{1}{t}\left[e^{2q_{1}} + e^{q_{2} - q_{1}}
+ \frac{1}{4}(-p_1 + p_2 + t)^2\right] .
\end{equation}
Making use of \eqref{eq1} and \eqref{eq4} leads to
\begin{equation}\label{eq7}
p_1 p_2 > e^{2q_{1}} + e^{q_{2} - q_{1}} +
\frac{1}{4}(p_1 + p_2 - t)^2 .
\end{equation}
It is clear that the signs of both r.h.s. of Eq. \eqref{eq5} and
Eq. \eqref{eq6} depend only on the sign of $t$.  It results from
\eqref{eq5}--\eqref{eq7} that the signs of both $p_1$ and $p_2$ are the same
for any value of $t\neq 0$.  The two sectors, $t>0$ and $t<0$, are dynamically
independent. In what follows we consider the case $t>0$, which implies $p_1 >
0$ and $p_2 > 0$.

The range of the variables $q_1$ and $q_2$ results from the physical
interpretation ascribed to them \cite{Czuchry:2012ad}. Since $0 <a < +\infty$
and $0 <b < +\infty$, we have $(q_1, q_2)\in \dR^2$.  Thus, the physical phase
space $\Pi$ consists of the two half planes:
\begin{equation}\label{kps2}
\Pi = \Pi_1 \times \Pi_2 :=
\{(q_1, p_1) \in \dR \times \dR_+\} \times
\{(q_2, p_2) \in \dR \times \dR_+\} \; ,
\end{equation}
where $\dR_+ := \{x \in \dR~|~x>0 \}$.

\section{Quantization}

Suppose we have reduced phase space Hamiltonian formulation of classical
dynamics of a gravitational system. It means dynamical constraints have been
resolved and the Hamiltonian is a generator of the dynamics.  By quantization we
mean (roughly speaking) a mapping of such Hamiltonian formulation into a quantum
system described in terms of quantum observables (including Hamiltonian)
represented by a set of self-adjoint operators acting in a Hilbert space. The
construction of the Hilbert space may make use some mathematical properties of
phase space like, e.g., symplectic structure, geometry or topology.  The quantum
Hamiltonian is used to define the Schr\"{o}dinger equation.  In what
follows we make specific the above procedure by using the affine coherent states
approach. Since the variables
parameterizing our phase space $\Pi$ are dimensionless, it is convenient to
assume $\hbar =1$ (unless otherwise specified) at the quantum level.

\subsection{Affine coherent states}

The Hilbert space $\mathcal{H}$ of the entire system consists of the Hilbert
spaces ${\mathcal{H}}_1$ and ${\mathcal{H}}_2$ corresponding to the phase spaces
$\Pi_1$ and $\Pi_2$, respectively. In the sequel the construction of
$\mathcal{H}_1$ is followed by merging of $\mathcal{H}_1$ and
$\mathcal{H}_2$.

As both half-planes $\Pi_1$ and $\Pi_2$ have the same mathematical structure,
the corresponding Hilbert spaces $\mathcal{H}_1$ and $\mathcal{H}_2$ are
identical so we first consider only one of them. In what follows we present the
formalism for $\Pi_1$ and $\mathcal{H}_1$ to be extended later to the entire
system.

\subsubsection{Affine coherent states for half-plane}

The phase space $\Pi_1$ may be identified with the affine group $\Group{G}_1
\equiv \Group{Aff}(\dR)$ by defining the multiplication law as follows
\begin{equation}\label{c1b}
(q^\prime, p^\prime)\cdot (q, p) = (p^\prime q+ q^\prime, p^\prime p ),
\end{equation}
with the unity $(0,1)$ and the inverse
\begin{equation}\label{c2b}
(q^\prime, p^\prime)^{-1} = (-\frac{q^\prime}{p^\prime}, \frac{1}{p^\prime}).
\end{equation}
The affine group has two, nontrivial, inequivalent irreducible unitary
representations \cite{Gel} and \cite{AK1,AK2}.  Both are realized in the Hilbert space
$\mathcal{H}_1=L^2(\dR_+, d\nu(x))$, where $d\nu(x)=dx/x$ is the invariant
measure\footnote{The general notion of
  invariant measure $dm(x)$ on the set $X$ in respect to the transformation $h:
  X \to X$ can be approximately defined as follows: for every function $f\colon
  X \to \dC$ the integral defined by this measure fulfils the invariance
  condition:
%
\[     \int_X dm(x) f(h(x)) = \int_X dm(x) f(x) \, .\]
%
This property is often written as: $dm(h(x))=dm(x)$.} on the multiplicative group $(\dR_+,\cdot)$. In what follows we choose a
different representation defined by the following action:
\begin{equation}\label{im1b}
U(q,p)\psi(x)= e^{i q x} \psi(px)\, ,
\end{equation}
where\footnote{We use Dirac's notation whenever we wish to deal with abstract
  vector, instead of functional representation of the vector.}  $|\psi\rangle
\in L^2(\dR_+, d\nu(x))$. Eq. \!\eqref{im1b} defines the representation as we
have
\begin{equation}\nonumber
U(q',p')[U(q,p,)\psi(x)]=U(q',p')[e^{iqx} \psi(px)]=
 e^{i(p'q+q')x} \psi(p'px) \, ,
\end{equation}
and on the other hand
\begin{equation}\nonumber
[U(q',p')U(q,p,)]\psi(x)= U(p'q+q',p'p) \psi(x) = e^{i(p'q+q')x} \psi(p'px)\, .
\end{equation}
This action is unitary in respect to the scalar product in
$L^2(\dR_+, d\nu(x))$:
\begin{eqnarray} \label{ActionUnitarity}
&& \int_0^\infty d\nu(x) [U(q,p) f_2(x)]^\star [U(q,p) f_1(x)]
= \int_0^\infty d\nu(x) [e^{iqx}f_2(px)]^\star [e^{iqx} f_1(px)] \nonumber \\
&& = \int_0^\infty d\nu(x) f_2(px)^\star f_1(px)=
\int_0^\infty d\nu(x) f_2(x)^\star f_1(x)\, .
\end{eqnarray}
The last equality results from the invariance of the measure
$d\nu(px)=d\nu(x)$.

The affine group is not the unimodular group. The left and right invariant
measures are given by
\begin{equation} \label{LRMeasuresAff}
d\mu_L(q,p)=dq\, \frac{dp}{p^2}
\quad \mbox{ and } \quad
d\mu_R(q,p)=dq\, \frac{dp}{p},
\end{equation}
respectively.

The left and right shifts of any  group $\Group{G}$ are defined
differently by different authors.  Here we adopt the definition from
\cite{Jin-Quan}:
\begin{equation} \label{leftRightShiftDef}
\mathcal{L}^L_h f(g)= f(h^{-1}g)
\quad \mbox{and} \quad
\mathcal{L}^R_h f(g)= f(gh^{-1})
\end{equation}
for a function  $f: \mathrm{G} \to \dC$  and all $g \in \Group{G}$.

For simplicity of notation, let us define integrals over the affine group
$\Group{G}_1 =\Group{Aff}(\dR)$ as:
\begin{equation}\label{HaarIntegrals}
\int_{\Group{G}_1} d\mu_L(q,p)= \frac{1}{2\pi}
\int_{-\infty}^{+\infty} dq \int_{0}^\infty \frac{dp}{p^2}
\quad \mbox{and} \quad
\int_{\Group{G}_1} d\mu_R(q,p)= \frac{1}{2\pi}
\int_{-\infty}^{+\infty} dq \int_{0}^\infty \frac{dp}{p} \, .
\end{equation}
In many formulae it is useful to use shorter notation for points in the phase
space $\xi \equiv (q,p)$ and identify them with elements of the affine group. In
this case the product (\ref{c1b}) is denoted as $\xi' \cdot \xi$.  Depending on
needs we will use both notations.

Fixing the normalized vector $\Ket{\Phi} \in L^2(\dR_+, d\nu(x))$, called the
{\it fiducial} vector, one can define a continuous family of {\it affine}
coherent states $\Ket{q,p} \in L^2(\dR_+, d\nu(x))$ as follows
\begin{equation}\label{im2}
\Ket{q,p} = U(q,p) \Ket{\Phi}.
\end{equation}
As we have two invariant measures, one can define two operators which
potentially can lead to the unity in the space $L^2(\dR_+, d\nu(x))$:
\begin{equation}\label{unityOneTwo}
B_L=\int_{\Group{G}_1} d\mu_L(q,p) \Ket{q,p}\Bra{q,p}
\quad \mbox{and} \quad
B_R=\int_{\Group{G}_1} d\mu_R(q,p) \Ket{q,p}\Bra{q,p} \, .
\end{equation}
Let us check which one is invariant under the action $U(q,p)$ of the affine
group:
\begin{equation} \label{BLInvU}
U(q',p') B_L U(q,p)^\dagger = \int_{\Group{G}_1} d\mu_L(q,p)
\Ket{p'q+q',p'p}\Bra{p'q+q',p'p} \, .
\end{equation}
One needs to replace the variables under integral:
\begin{eqnarray} \label{BLInvU2}
&& \tilde{q}=p'q+q' \quad \mbox{and} \quad \tilde{p}=p'p \\
&& q=\frac{1}{p'}(\tilde{q}-q') \quad \mbox{and} \quad p=\frac{\tilde{p}}{p'}.
\end{eqnarray}
Calculating the Jacobian
$\frac{\partial(q,p)}{\partial(\tilde{q},\tilde{p})}=\frac{1}{(p')^2}$ one gets
\begin{equation} \label{BLInvU3}
d\mu_L(q,p)=\frac{1}{p^2} \frac{1}{(p')^2} d\tilde{q}d\tilde{p}=
\frac{1}{\tilde{p}^2} d\tilde{q}d\tilde{p}= d\mu_L(\tilde{q},\tilde{p})\, .
\end{equation}
The last result proves that
\begin{equation} \label{BLInvU}
U(q',p') B_L U(q,p)^\dagger = \int_{\Group{G}_1}
d\mu_L(\tilde{q},\tilde{p}) \Ket{\tilde{q},\tilde{p}}\Bra{\tilde{q},\tilde{p}}
= B_L \, .
\end{equation}
This also means that $B_R$ is not invariant under the action $U(q,p)$.

The {\it irreducibility} of the representation, used to define the coherent
states \eqref{im2}, enables making use of Schur's lemma \cite{BR}, which leads
to the resolution of the unity in $L^2(\dR_+, d\nu(x))$:
\begin{equation}\label{im4}
\int_{\Group{G}_1}  d\mu_L(q,p) \Ket{q,p}\Bra{q,p} = A_\Phi\;\I \; ,
\end{equation}
where the constant $A_\Phi$ can be calculated using any arbitrary, normalized
vector $\Ket{f} \in L^2(\dR_+, d\nu(x))$:
\begin{equation}\label{im3b}
A_\Phi = \int_{\Group{G}_1}d\mu_L(q,p)\,
\BraKet{f}{q,p}\BraKet{q,p}{f} \, .
\end{equation}
This formula can be calculated by making use of invariance of the measure:
\begin{eqnarray}\label{im3b2}
&& A_\Phi = \int_{\Group{G}_1}  d\mu_L(q,p) \nonumber \\
&& \int_0^\infty d\nu(x')\int_0^\infty d\nu(x)
(f(x')^\star e^{iqx'}\Phi(px'))(e^{-iqx}\Phi(px)^\star f(x)) \nonumber \\
&& = \int_0^\infty \frac{dx'}{x'}\int_0^\infty \frac{dx}{x}
\int_0^\infty \frac{dp}{p^2}
\left[\frac{1}{2\pi}\int_{-\infty}^{+\infty} dq e^{iq(x'-x)}\right]
f(x')^\star f(x) \Phi(px')\Phi(px)^\star \nonumber \\
&& = \int_0^\infty \frac{dx}{x^2} |f(x)|^2
\int_0^\infty \frac{dp}{p^2} |\Phi(px)|^2   \nonumber \\
&& = \left(\int_0^\infty \frac{dx}{x} |f(x)|^2\right)
\left(\int_0^\infty \frac{dp}{p^2} |\Phi(p)|^2 \right)
=\int_0^\infty \frac{dp}{p^2} |\Phi(p)|^2
\end{eqnarray}
because  $\BraKet{f}{f}=1$. Thus, the normalization constant is dependent
on the fiducial vector.

\subsubsection{Structure of the fiducial vector}

The problem which influences the structure of quantum state space is a
possible degeneration of the space due to specific structure of the fiducial
vector. In the case of quantum states the vectors which differ by a phase factor
represent the same quantum state.
Thus, let us consider the states satisfying the above condition for physically
equivalent state vectors \cite{AP}:
\begin{equation}
 \label{FidVectPhase}
 U(\tilde{q},\tilde{p}) \Phi(x)=
 e^{i\beta(\tilde{q},\tilde{p})} \Phi(x),~~
 \mbox{ where }~~~~
 \beta(\tilde{q},\tilde{p}) \in \dR.
 \end{equation}
 The phase space points $\tilde{\xi}=(\tilde{q},\tilde{p})$ treated as elements
 of the affine group $\Group{Aff}(\dR)$ forms its subgroup $\Group{G}_\Phi$.
 The left-hand side of Eq.~\eqref{FidVectPhase} can be rewritten as:
\begin{equation} \label{FidVectPhase2}
e^{i\tilde{q}x}\Phi(\tilde{p})=
e^{i \beta(\tilde{q},\tilde{p})} \Phi(x)\, .
 \end{equation}

 If the generalized stationary group $\Group{G}_\Phi$ of the fiducial vector
 $\Phi$ is a nontrivial group, then the phase space points $(q',p')$ and
 $(\tilde{q},\tilde{p}) \cdot (q',p') = (\tilde{p}q',\tilde{p}p')$ are
 represented by the same state vector $U(q',p') \Phi(x)$, for all
 transformations $(\tilde{q},\tilde{p}) \in \Group{G}_\Phi$. This is due to the
 equality
\begin{equation} \label{FidVectPhas3}
U(q',p') \Phi(x)= U((\tilde{q},\tilde{p}) \cdot (q',p')) \Phi(x).
\end{equation}
In this case, to have a unique relation between phase space and the quantum
states, the phase space has to be restricted to the quotient structure
$\Group{Aff}(\dR)/ \Group{G}_\phi$. From the physical point of view, in most
cases, this is an undesired property.

How to construct the fiducial vector to have $\Group{G}_\Phi=\{e_G\}$, where
$e_G$ is the unit element in this group?

It is seen that Eq.~\eqref{FidVectPhase2} cannot be fulfilled for $\tilde{q}
\neq 0$, independently of chosen fiducial vector. This suggests that the
generalized stationary group $\Group{G}_\Phi$ is parameterized only by the
momenta $(0,\tilde{p})$, i.e. it has to be a subgroup of the multiplicative
group of positive real numbers, $\Group{G}_\Phi \subseteq (\dR_+,\cdot)$.

On the other hand, Eq.~\eqref{FidVectPhase2} implies that
$|\Phi(\tilde{p} x)|= |\Phi(x)|$ for all $(0,\tilde{p}) \in
\Group{G}_\Phi$. In addition, for the fiducial vectors
$\Phi(x)=|\Phi(x)|e^{i\gamma(x)}$ the phases of these complex functions are
bounded by $0\leq \gamma(x) < 2\pi$. Due to Eq.~\eqref{FidVectPhase2} the phases
$\gamma(x)$ and $\beta(0,\tilde{p})$ have to fulfil the following condition
$\gamma(\tilde{p}x)-\gamma(x)=\beta(0,\tilde{p})$. One of the
solutions to this equation is the logarithmic function $\gamma(x)=  \ln(x)$.

In what follows, to have the unique representation of the phase space as a group
manifold of the affine group, we require the generalized stationary group to be
the group consisted only of the unit element. This can be achieved by the
appropriate choice of the fiducial vector.

The unit operator (\ref{im4}) depends explicitly on the fiducial vector
\begin{equation}
\label{UnitOperPhi}
\I[\Phi]= \frac{1}{A_\Phi}
\int_{\Group{G}_1} d\mu_L(\xi) U(\xi)\Ket{\Phi}\Bra{\Phi}U(\xi)^\dagger
\; ,
\end{equation}
This suggests that the most natural transformation of vectors from the
representation given by the fiducial vector $\Ket{\Phi}$ to the representation
given by another fiducial vector $\Ket{\Phi'}$ can be constructed as the
product of two unit operators $\I[\Phi']\I[\Phi]$.

Let us consider an arbitrary vector $\Ket{\Psi}$ from the space $L^2(\dR_+,
d\nu(x))$ and its representation in the space spanned with a help of the
fiducial vector $\Ket{\Phi}$:
\begin{equation}
\label{VectorRepFidVectPhi}
\Ket{\Psi}=\I[\Phi]\Ket{\Psi}= \frac{1}{A_\Phi}
\int_{\Group{G}_1} d\mu_L(\xi)
U(\xi)\Ket{\Phi}\Bra{\Phi}U(\xi)^\dagger\Ket{\Psi}
\end{equation}
The same vector can be represented in terms of another fiducial vector
$\Ket{\Phi'}$:
\begin{equation}
\label{VectorRepFidVectTildePhi}
\Ket{\Psi}=\I{[\Phi'] \Ket{\Psi}}= \frac{1}{A_{\Phi'}}
\int_{\Group{G}_0} d\mu_L(\xi)
U(\xi)\Ket{\Phi'} \Bra{\Phi'} U(\xi)^\dagger\Ket{\Psi}
\end{equation}
However, one can transform the vector (\ref{VectorRepFidVectPhi}) into the
vector (\ref{VectorRepFidVectTildePhi}) using the product of two unit operators:
\begin{eqnarray}
\label{TransRepFVPhiFVTildePhi}
&& \Ket{\Psi}=\I [\Phi']\I[\Phi]\Ket{\Psi}= \frac{1}{A_{\Phi'}}\frac{1}{A_\Phi}
\int_{\Group{G}_1} d\mu_L(q,p)
\int_{\Group{G}_1} d\mu_L(q',p')
U(\xi)\Ket{\Phi'} \nonumber \\
&& \Bra{\Phi'} U(\xi^{-1} \cdot \xi') \Ket{\Phi}
   \Bra{\Phi} U(\xi')^\dagger \Ket{\Psi}
\end{eqnarray}
In this way, we see that a choice of the fiducial vector is formally
irrelevant. However, as we will see later a relation between the classical model
and its quantum realization depends on this choice. The sets of affine coherent
states generated from different fiducial vectors may be not unitarily
equivalent, but lead in each case to acceptable affine representations of the
Hilbert space \cite{JRK}.

\subsubsection{Phase space and quantum state spaces}

The quantization procedure requires understanding the relations among the
classical phase space and quantum states space. We have three spaces to be
considered:
\begin{itemize}
\item The phase space $\Pi$, which consists of two half-planes $\Pi_1$ and
  $\Pi_2$ defined by \eqref{kps2}. It is the background for the classical
  dynamics\footnote{For simplicity we consider here only one half-plane, but the
    results can be easily extended to $\Pi$.}.
\item The carrier spaces $ \mathcal{H}_1 :=L^2(\dR_+,
  d\nu(x))$ of the unitary representation $U(q,p)$, with the scalar product
  defined as
\begin{equation}
\label{ScalarProdKX}
\BraKet{\psi_2}{\psi_1}
= \int_0^\infty d\nu(x) \psi_2^\star (x) \psi_1(x)\, .
\end{equation}
\item The  space of square integrable functions on the
  affine group $\mathcal{K}_G = L^2(\Group{Aff}(\dR), d\mu_L(q,p))$.
  The scalar product is defined as follows
\begin{equation}
\label{ScalarProdKG}
\BraKet{\psi_{G2}}{\psi_{G1}}_{G}
= \frac{1}{A_\phi}\int_{\Group{Aff}} d\mu_L(q,p)
\psi_{G2}^\star(q,p) \psi_{G1}(q,p)\, ,
\end{equation}
where $\psi_G(q,p):=\BraKet{q,p}{\psi}=\Bra{\Phi}U(q,p)^\dagger\Ket{\psi}$ with
$\Ket{\psi} \in \mathcal{H}_1$. The Hilbert space $\mathcal{K}_G $ is defined to
be the completion in the norm induced by \eqref{ScalarProdKG} of the span of the
$\psi_G$ functions.
\end{itemize}
We show below that the spaces $\mathcal{H}_1$  and $\mathcal{K}_G$
are unitary isomorphic.  First, one needs to check that the functions $ \psi_G
\in \mathcal{K}_G$ are square integrable function belonging to
$L^2(\Group{Aff}(\dR), d\mu_L(q,p))$. Using the decomposition of unity we get
\begin{equation}
\label{L2GSquareIntegFun}
\frac{1}{A_\Phi} \int_{\Group{G}_1} d\mu_L(q,p)|\BraKet{q,p}{\psi}|^2
< \BraKet{\psi}{\psi}_{\mathcal{H}_1} < \infty.
\end{equation}
The definition of the space $\mathcal{K}_G$ shows that for every
$\psi_1,\psi_2 \in \mathcal{H}_l$ we have the corresponding functions
$\psi_{G1},\psi_{G2}$ for which the scalar products are equal (unitarity of the
transformation between both spaces): $\BraKet{\psi_2}{\psi_1}_{\mathcal{H}_1}=
\BraKet{\psi_{G1}}{\psi_{G2}}_{\mathcal{K}_G}$.

Let us now denote by $\Ket{e_n}$ the orthonormal basis in $\mathcal{H}_1$ (see
App. \!\ref{basis}). The corresponding functions $e_{Gn}(q,p)
=\BraKet{q,p}{e_n}$ furnish the orthonormal set:
\begin{eqnarray}
\label{L2GBasis}
&& \BraKet{e_{Gn}}{e_{Gm}}_{\mathcal{K}_G}
=\frac{1}{A_\Phi} \int_{\Group{G}_1} d\mu_L(q,p)
e_{Gn}^\star (q,p)  e_{Gm}(q,p) \nonumber \\
&& = \frac{1}{A_\Phi} \int_{\Group{G}_1} d\mu_L(q,p)
\BraKet{e_n}{q,p}  \BraKet{q,p}{e_m}
= \BraKet{e_n}{e_m}=\delta_{nm} \, .
\end{eqnarray}
It is obvious that the vectors $\Ket{e_{Gn}}$ define the orthonormal basis in
the space $\mathcal{K}_G$. For every vector $\Ket{\psi} \in \mathcal{H}_1$
\begin{equation}
\label{BasisKX}
\Ket{\psi}= \sum_n \BraKet{e_n}{\psi} \Ket{e_n}.
\end{equation}
Closing both sides of the above equation with $\Bra{q,p}$ gives the unique
decomposition of the vector $\psi_G(q,p) \equiv \BraKet{q,p}{\psi} \in
\mathcal{K}_X $ in the basis $\Ket{e_{Gn}}_G$:
\begin{equation}
\label{BasisKX2}
\psi_G(q,p) \equiv \BraKet{q,p}{\psi}
= \sum_n \BraKet{e_n}{\psi} \BraKet{q,p}{e_n}
= \sum_n \BraKet{e_n}{\psi} \Ket{e_{Gn}}_G.
\end{equation}
Note that the vector $\Ket{\psi} \in \mathcal{H}_1$ and the vector $\Ket{\psi_G}
\in \mathcal{K}_G$ have the same expansion coefficients in the corresponding
bases. This define the unitary isomorphism between both spaces. It means that we
can work either with the quantum state space represented by the space
$\mathcal{H}_1$ or $\mathcal{K}_G$.

\subsubsection{Affine coherent states for the entire system}

The phase space $\Pi$ of our classical system has the structure of the Cartesian
product of two partial phase spaces $\Pi_1$ and $\Pi_2$: $\Pi=\Pi_1 \times
\Pi_2$.  The partial phase spaces $\Pi_l$, where $l=1,2$, are identified with
the corresponding affine groups which we denote by
$\Group{G}_l=\Group{Aff}_l(\dR)$. The simple product of both affine groups
$\Group{G}_{\Pi}= (\Group{G}_1=\Group{Aff}_1(\dR)) \times
(\Group{G}_2=\Group{Aff}_2(\dR)) $ can be identified with the whole phase space
$\Pi$:
\begin{equation}
\label{AffActPi}
(\xi_1,\xi_2) \to \Ket{\xi_1,\xi_2 } = U(\xi_1,\xi_2)\Ket{\Phi}
:= U_1(\xi_1) \otimes U_2(\xi_2) \Ket{\Phi},
\end{equation}
where $\xi_l=(q_l,p_l)$, $l=1,2,$ the fiducial vector $\Ket{\Phi}$ belongs to
the simple product of two Hilbert spaces $(\mathcal{H}_1=L^2(\dR_+,d\nu(x_1)))
\times (\mathcal{H}_2=L^2(\dR_+,d\nu(x_2))) = L^2(\dR_+ \times
\dR_+,d\nu(x_1,x_2))$, and where the measure
$d\nu(x_1,x_2)=d\nu(x_1)d\nu(x_2)$. The scalar product in $\mathcal{H}=L^2(\dR_+
\times \dR_+,d\nu(x_1,x_2))$ reads
\begin{equation}
\label{ScalarProdL2Whole}
\BraKet{\psi_2}{\psi_1}= \int_0^\infty d\nu(x_1) \int_0^\infty d\nu(x_2)
\psi_1(x_1,x_2)^\star \psi_2(x_1,x_2) \, .
\end{equation}
The fiducial vector $\Phi(x_1,x_2)$ is constructed as a product of two fiducial
vectors $\Phi(x_1,x_2)= \Phi_1(x_1)\Phi_2(x_2)$ generating the appropriate
quantum partners for the phase spaces $\Pi_1$ and $\Pi_2$. The fiducial vector
of this type does not add any correlations between both partial phase
spaces.  A nonseparable form of $\Phi(x_1,x_2)$ might lead to reducible
representation of $G_\Pi$ in which case Schur's lemma could not be applied to
get the resolution of unity in $\mathcal{H}$.

Let us denote by $\UnitOp_{12}$ the linear extension of the tensor
product $\UnitOp_{1} \otimes \UnitOp_{2}$, where the unit operators in
$\mathcal{H}_k$ are expressed in terms of the appropriate coherent states
\begin{equation}
\label{UnitOperSpacel}
\UnitOp_{k}= \frac{1}{A_{\Phi_k}}
\int_{\Group{G}_k}  d\mu_L(\xi_k)
U_k(\xi_k)\Ket{\Phi_k}\Bra{\Phi_k}U_k(\xi_k)^\dagger,~~~~k = 1,2.
\end{equation}
Let us consider the orthonormal basis $\{e^{(1)}_n(x_1) \otimes
e^{(2)}_n(x_2)\}$ in the Hilbert space $\mathcal{H}$ and an arbitrary vector
$\Psi(x_1,x_2)= \sum_{nm} a_{nm} e^{(1)}_n(x_1) \otimes e^{(2)}_m(x_2)$
belonging to this space (where the basis $e_n(x)$ is defined in
App. \ref{basis}).  Acting on this vector with the operator $\UnitOp_{12}$ one
gets:
\begin{eqnarray}
\label{ActUnitOp12Vect}
\UnitOp_{12} \Psi(x_1,x_2)= \sum_{nm} a_{nm} (\UnitOp_{1}e^{(1)}_n(x_1)) \otimes
(\UnitOp_{2} e^{(2)}_m(x_2))= \Psi(x_1,x_2).
\end{eqnarray}
The operator $\UnitOp_{12}$ is identical with the unit operator $\UnitOp$ on the
space $\mathcal{H}$.

The explicit form of the action of the group
$\Group{G}_{\Pi}$ on the vector $\Psi(x_1,x_2)$ reads:
\begin{eqnarray}
\label{ActGPiH}
&& U(q_1,p_1,q_2,p_2)\Psi(x_1,x_2)=
\sum_{nm} a_{nm} \{U_1(q_1,p_1)e^{(1)}_n(x_1)\}
\otimes \{U(q_2,p_2) e^{(2)}_m(x_2)\} \nonumber \\
&& =\sum_{nm} a_{nm} \{e^{iqx_1} e^{(1)}_n(p_1 x_1)\}
\otimes \{e^{iqx_2} e^{(2)}_m(p_2x_2)\}
=  e^{iq_1x_1} e^{iq_2x_2} \sum_{nm} a_{nm} e^{(1)}_n(p_1 x_1)
\otimes e^{(2)}_m(p_2x_2) \nonumber \\
&& = e^{iq_1x_1} e^{iq_2x_2} \Psi(p_1 x_1,p_2 x_2)\, .
\end{eqnarray}

\subsection{Quantum observables}

Making use of the resolution of the identity \eqref{im4}, we define the
quantization of a classical observable $f$ on a half-plane as follows \cite{Ber}
\begin{equation}\label{im8}
  \mathcal{F} \ni f \longrightarrow  \hat{f} :=
\frac{1}{A_\Phi}\int_{\Group{G}_1} d\mu_L(q,p)
\Ket{q,p} f(q,p) \Bra{q,p}  \in \mathcal{A} \, ,
\end{equation}
where $\mathcal{F}$ is a vector space of real continuous functions on a phase
space, and $\mathcal{A}$ is a vector space of operators (quantum observables)
acting in the Hilbert space $\mathcal{H}_1=L^2(\dR_+, d\nu(x))$. It is clear
that \eqref{im8} defines a linear mapping and the observable $\hat{f}$ is a {\it
  symmetric} (Hermitian) operator. Let us evaluate the norm of the operator
$\hat{f}$:
\begin{equation}
\label{OperNormQuantOper}
\Vert\hat{f}\Vert \leq
\frac{1}{A_\Phi}\int_{\Group{G}_1} d\mu_L (q,p)
|f(q,p)|  \Vert \Ket{q,p}\Bra{q,p} \Vert
\leq \frac{1}{A_\Phi}\int_{\Group{G}_1} d\mu_L(q,p) |f(q,p)| \, .
\end{equation}
This implies that, if the classical function $f$ belongs to the space of
integrable functions $L^1(\Group{Aff(\dR)}, d\mu_L(q,p))$, the operator
$\hat{f}$ is bounded so it is a {\it self-adjoint} operator.  Otherwise, it is
defined on a dense subspace of $L^2(\dR_+, d\nu(x))$, and its possible
self-adjointness becomes an open problem as symmetricity does not assure
self-adjointness, and further examination is required \cite{Reed}.  The
quantization \eqref{im8} can be applied to any type of observables including
non-polynomial ones, which is of primary importance for us due to the functional
form of the Hamiltonian \eqref{sim3}.

It is not difficult to show that the mapping \eqref{im8} is
covariant in the sense that one has
\begin{equation}\label{cov}
  U(\xi_0) \hat{f} U^\dag (\xi_0) =
\frac{1}{A_\Phi}\int_{\Group{G}_1} d\mu_L(\xi)
\Ket{\xi} f(\xi_0^{-1}\cdot \xi) \Bra{\xi}
=  \widehat{\mathcal{L}^L_{\xi_0}f} \, ,
\end{equation}
where $\mathcal{L}^L_{\xi_0} f(\xi) = f(\xi_0^{-1}\cdot \xi)$ is the left shift
operation (\ref{leftRightShiftDef}) and $\xi_0^{-1}\cdot \xi=
(q_0,p_0)^{-1}\cdot (q,p) = (\frac{q-q_0}{p_0},\frac{p}{p_0})$.

The mapping \eqref{im8} extended to the Hilbert space $\mathcal{H}=L^2(\dR_+
\times \dR_+,d\nu(x_1,x_2))$ of the entire system and applied to an observable
$\hat{f}$ reads
\begin{equation}\label{QuantObservPi}
 \hat{f}(t)= \frac{1}{A_{\Phi_1} A_{\Phi_2}}\int_{G_\Pi}  d\mu_L(\xi_1,\xi_2)
  \Ket{\xi_1,\xi_2}f(\xi_1,\xi_2)\Bra{\xi_1,\xi_2}\, ,
\end{equation}
where $ d\mu_L(\xi_1,\xi_2) := d\mu_L(q_1,p_1)d\mu_L(q_2,p_2)$.

\section{Quantum dynamics}

The mapping \eqref{QuantObservPi} applied to the classical Hamiltonian
\eqref{sim3} reads
\begin{equation}\label{Qd1}
 \hat{H}(t)= \frac{1}{A_{\Phi_1} A_{\Phi_2}}\int_{G_\Pi}  d\mu_L(\xi_1,\xi_2)
  \Ket{\xi_1,\xi_2}H(\xi_1,\xi_2,t)\Bra{\xi_1,\xi_2}\, ,
\end{equation}
where $t$ is an evolution parameter of the classical level.

Suppose that $\hat{H}$ is bounded on $\mathcal{H}$ so it is self-adjoint on
$\mathcal{H}$. Therefore, we can define the quantum evolution using the
Schr\"{o}dinger equation as follows
\begin{equation}\label{Qd2}
  i  \frac{\partial}{\partial \tau}|\psi (\tau) \rangle
= \hat{H}(t) |\psi (\tau) \rangle \; ,
\end{equation}
where $|\psi \rangle \in \mathcal{H}$, and where $\tau$ is an evolution
parameter at the quantum level.

In general, the parameters $t$ and $\tau$ are quite different. To get the
consistency between the classical and quantum levels we postulate that $t =
\tau$, which defines the {\it time} variable at both levels.

\section{The probability measure}

Let us define the following probability measure on the phase space:
\begin{equation}\label{pro1}
  \Pi \supset\Omega \rightarrow \hat{M}(\Omega):=
\frac{1}{A_{\Phi_1}A_{\Phi_2}}\int_\Omega  d\mu_L(\xi_1)d\mu_L(\xi_2)
\Ket{\xi_1, \xi_2}\Bra{\xi_1, \xi_2}\, ,
\end{equation}
where $\Ket{\Phi_1} \in \mathcal{H}_1$ and $\Ket{\Phi_2} \in
\mathcal{H}_2$ are the fiducial vectors. This probability measure
represents the observable: ``the system is in the region $\Omega$ of the phase
space $\Pi$''.

In what follows we take $\Phi_1(x) = \Phi_2(x) = \Phi(x)$ because the
Hilbert spaces $\mathcal{H}_1$ and $\mathcal{H}_2$ are identical, and any two
fiducial vectors can be linked by the unitary map of the form \eqref{im1b}.

The measure \eqref{pro1} is of the POV type \cite{Busch1996} as it is
\begin{itemize}\label{pro2}
\item positive,  $\hat{M}(\Omega)\geq 0 $, $\forall \,\Omega \subset \Pi$;
\item additive,  $\hat{M}(\bigcup_k\Omega_k) = \sum_k \hat{M}(\Omega_k)~ $
  if $~\Omega_i \bigcap \Omega_j =\varnothing,~~ i\neq j$ ;
\item normalized,  ${\hat{M}(\Pi)} = \id$ .
\end{itemize}
The volume $V(\Omega)$ of $\Omega$ is defined as
\begin{equation}\label{pro3}
 V(\Omega) =  \int_\Omega  d\mu_L(q_1,p_1) d\mu_L(q_2,p_2)\, .
\end{equation}
One needs to remember the factor $1/(2\pi)^2$ hidden in this integral, see
\ref{HaarIntegrals}.

Now, suppose that $\Omega$ is a small neighbourhood of $(\xi_1,\xi_2) :=
(q_1,p_1, q_2,p_2)$.  We define the {\it probability density} measure operator
$ \hat{M}(\xi_1,\xi_2)$, which determines the probability that the quantum
system is at the state $\Ket{\xi_1, \xi_2}$  of the phase space $\Pi$. It is defined
as follows
\begin{equation}\label{pro4}
  \hat{M}(\xi) =
\lim_{ V(\Omega)\rightarrow 0}\frac{\hat{M}(\Omega)}{ V(\Omega)} \,
\mbox{ where } (\xi_1,\xi_2) \in \Omega.
\end{equation}
The r.h.s. of \eqref{pro4} is well defined due to the mean value theorem
applied to the measure \eqref{pro1}.  This  probability density
operator can be rewritten as
\begin{equation}\label{pro5}
 \hat{M}(\xi_1,\xi_2)= \frac{1}{A_{\Phi}^2}\,
\Ket{\xi_1,\xi_2}\Bra{\xi_1,\xi_2}  \, .
\end{equation}
Let $f : \Pi \rightarrow \dR$ be a classical observable and $\Ket{\psi}$ a
quantum state of our system.  The expectation value of the corresponding quantum
operator $\hat{f}$ is given by the standard expression
\begin{equation}\label{pro6}
  \langle \hat{f}\rangle = \Bra{\psi}\hat{f} \Ket{\psi}
=  \frac{1}{ A_{\Phi}^2} \int_{G_\Pi} d \mu_L (\xi_1,\xi_2)
\BraKet{\psi}{\xi_1,\xi_2} f(\xi_1,\xi_2) \BraKet{\xi_1,\xi_2}{\psi} \, ,
\end{equation}
where $d \mu_L (\xi_1,\xi_2) := d \mu_L (q_1,p_1) d \mu_L (q_2, p_2)$.

Let $\Pi = \cup_k \Omega_k$, with $\Omega_i \cap \Omega_j= \varnothing $ if $i
\neq j$. The probability density that $(\xi_{1k},\xi_{2k}) \in \Omega_k$ is
given by
\begin{equation}\label{pro7}
  p\big((\xi_{1k},\xi_{2k}) \in \Omega_k):=
\Bra{\psi}\hat{M}(\Omega_k\big) \Ket{\psi} \, .
\end{equation}
The consistency between \eqref{pro6} and \eqref{pro7} occurs if one has
\begin{equation}\label{pro8}
\langle \hat{f}\rangle \approx
\sum_k f (\xi_{1k},\xi_{2k})  p\big((\xi_{1k},\xi_{2k}) \in \Omega_k\big).
\end{equation}
One can show that \eqref{pro8} is satisfied due to the mean value theorem
applied to the measure \eqref{pro1}.  In the limit $V(\Omega_k) \rightarrow 0$,
such that $\Pi = \cup_k \Omega_k$  Eq. \!\eqref{pro8} is exact.

\section{Prospects}

Recently, special structures have been found in the evolution of gravitational
systems. They are named spikes \cite{Czuchry:2016rlo} and wiggles
\cite{Piechocki:2016okx}.

To compare the classical and quantum dynamics of the wiggles, one needs to
compare the classical solutions $\zeta(t) =
(\zeta_1(t),\zeta_2(t),\zeta_3(t),\zeta_4(t)) := (q_1(t),p_1(t),q_2(t),p_2(t))$
of the Hamilton equations \!\eqref{ff1}--\eqref{ff4} with the average values of
the corresponding quantum operators:
\begin{equation}
\label{PositionOperator}
\hat{\zeta}_k
=  \frac{1}{ A_{\Phi}^2} \int_{\Group{G}_\Pi} d\mu_L (\zeta)
\Ket{\zeta}\zeta_k \Bra{\zeta}
= \int_{\Group{G}_\Pi} d\mu_L (\zeta) \zeta_k \hat{M}(\zeta) \, .
\end{equation}
The expectation value is given by
\begin{equation}
\label{AverPositionOperator}
\Aver{\hat{\zeta}_k}_{\psi(t)} = \Bra{\psi(t)}\hat{\zeta}_k \Ket{\psi(t)}
=  \frac{1}{ A_{\Phi}^2} \int_{\Group{G}_\Pi}  d\mu_L (\zeta)
\BraKet{\psi(t)}{\zeta} \zeta_k \Bra{\zeta}{\psi(t)} \, .
\end{equation}
This average is calculated in the state $\Ket{\psi(t)}$ evolving according to
the Schr\"{o}dinger equation \eqref{Qd2} with the initial condition
corresponding to the initial condition used in solution of the Hamilton
equations.  If the classical initial condition is given by
$\zeta_k(t_0)=\zeta^{(0)}_k$, then the corresponding initial condition for the
Schr\"odinger equation is defined to be
\begin{equation}\label{qcon}
\Bra{\psi(t_0)} \hat{\zeta}_k
\Ket{\psi{(t_0)}}=\zeta^{(0)}_k.
\end{equation}
In such a case the classical trajectories $\zeta_k(t)$ and the quantum
``trajectories'' $\Aver{\hat{\zeta}_k}_{\psi(t)}$ are treated on the same
footing.

The wiggles are parametric curves, $\zeta : \R \supset [t_1,t_2] \rightarrow
\Pi \subset\R^4$, in the physical phase space $\Pi$. The {\it observables}
invariant with respect to reparametrization of a curve can be chosen to be
\cite{WK}:
\begin{itemize}
  \item the length $s[\zeta]$ of a curve (global observable),
  \item the generalized curvatures $\{\chi_1(t), \chi_2(t), \chi_3(t) \}$ (local observables) of a curve defined by the Frenet vectors.
\end{itemize}
The curvatures of these curves depend on the location of the initial conditions in the physical phase spaced $\Pi$ specifying the
dynamics, and on the stage of the evolution of the system.

Both the length and the curvatures can be determined for the quantum  ``curves'' $\Aver{\hat{\zeta}_k}_{\psi(t)}$ as well. Thus,
the comparison of  the observables for the classical and quantum curves will enable finding the influence of the quantization
on the classical wiggles.

Does quantization suppress these structures, leave them almost unchanged, or turns  them into some quantum structures?
Our next paper will be devoted to the examination of this issue.

\acknowledgments We would like  to thank Katarzyna G\'{o}rska for the derivation of Eqs. \!\eqref{eq5}--\eqref{eq7}, and
John Klauder for helpful correspondence.

\appendix

\section{Regularized Hamiltonian}
\label{regular}
The Hamiltonian \eqref{sim3} is restricted to the region of $\Pi$ corresponding
to $F>0$. We extend it to the whole phase space by introducing an additional
functional coefficient which vanishes for $F \leq 0$. It has the form
\begin{equation}
\label{Hnew1}
H_{new} := \theta \left[ p_1 p_2 - e^{2q_{1}} - e^{q_{2} - q_{1}}
- \frac{1}{4}(p_1 + p_2 - t)^2  \right] H(q_1,q_2;p_1,p_2;t)\, ,
\end{equation}
where $H(q_1,q_2;p_1,p_2;t)$ is the original Hamiltonian \eqref{sim3}, while
$\theta$ stands for the Heaviside step function.  However, so defined new
Hamiltonian is not differentiable at $F=0$. To get a better behaviour we adopt
the following ``smooth'' version of $\theta$:
\begin{equation}
\label{thetaeps}
\theta_{\epsilon}(x) :=  \begin{cases}
0 \, : x \, \leq 0, \\
\frac{1}{2} (1-\cos(\frac{\pi x }{\epsilon})) \, : 0 < x \leq \epsilon, \\
1 : x> \epsilon.
\end{cases}
\end{equation}
Here, $\epsilon$ is a small parameter $0 < \epsilon \ll 1$. Using
\eqref{thetaeps} we rewrite \eqref{Hnew1}:
\begin{equation}
\label{Hnew2}
H_{\epsilon}(q_1,q_2;p_1,p_2;t) :=
\theta_{\epsilon} \left[ p_1 p_2 - e^{2q_{1}} - e^{q_{2} - q_{1}}
-\frac{1}{4}(p_1 + p_2 - t)^2  \right] H(q_1,q_2;p_1,p_2;t) \, .
\end{equation}
The Hamiltonian $H_{\epsilon}$ is well defined and differentiable in the whole
phase space. Indeed, for any $(q_1,q_2;p_1,p_2;t)$ corresponding to $F>0$ one
finds $H_{\epsilon}(q_1,q_2;p_1,p_2;t) \simeq H(q_1,q_2;p_1,p_2;t)$.  The
smaller $\epsilon$ is, the better approximation we get.  Also the function
\eqref{thetaeps} guaranties that the Hamiltonian \eqref{Hnew2} vanishes
smoothly in the limit $F \rightarrow 0$.
Taking the limit $\epsilon \rightarrow 0^+$ restores to the original
expression \eqref{sim3}.

\section{Alternative affine coherent states for half-plane}
\label{alternative}

The phase space $\Pi_1$ may be identified with the affine group
$\Group{Aff}(\dR)$ by defining the multiplication law as follows
\begin{equation}\label{c1}
 (q^\prime, p^\prime)\cdot (q, p) =(\frac{q}{p^\prime}+ q^\prime, p^\prime p ),
\end{equation}
with the unity $(0,1)$ and the inverse
\begin{equation}\label{c2}
(q^\prime, p^\prime)^{-1} = (-q^\prime p^\prime, \frac{1}{p^\prime}).
\end{equation}
The affine group has two, nontrivial, inequivalent irreducible unitary
representations \cite{Gel} and \cite{AK1,AK2}.  Both are realized in the Hilbert space
$L^2(\dR_+, d\nu(x))$, where $d\nu(x)=dx/x$ is the invariant measure on
the multiplicative group $(\dR_+,\cdot)$. In what follows we choose the one
defined by
\begin{equation}\label{im1}
 U(q,p)\psi(x)= e^{i q x} \psi(x/p)\, ,
\end{equation}
where $|\psi\rangle \in L^2(\dR_+, d\nu(x))$.

For simplicity of notation, let us define integrals over the affine group
$\Group{Aff}(\dR)$ as follows:
\begin{eqnarray}
&&\int_{\Group{ Aff}(\dR)} d\mu_L(q,p)= \frac{1}{2\pi}
\int_{-\infty}^{+\infty} dq \int_{0}^\infty \frac{dp}{p^2} \, ,\\
&&\int_{\Group{Aff}(\dR)} d\mu_R(q,p)= \frac{1}{2\pi}
\int_{-\infty}^{+\infty} dq \int_{0}^\infty \frac{dp}{p} \, ,\\
&&\int_{\Group{Aff}(\dR)} d\mu_U(q,p)= \frac{1}{2\pi}
\int_{-\infty}^{+\infty} dq \int_{0}^\infty dp \, \rho(q,p) \, .
\end{eqnarray}
The last one is intended to be used as invariant measure in respect
with the action $U(q,p)$.

Fixing the normalized vector $\Ket{\phi} \in L^2(\dR_+, d\nu(x))$, called the
{\it fiducial} vector, one can define a continuous family of {\it affine}
coherent states $\Ket{q,p} \in L^2(\dR_+, d\nu(x))$ as follows
\begin{equation}\label{im22}
\Ket{q,p} = U(q,p) \Ket{\phi}.
\end{equation}
As we have three  measures,  one can define three operators which
potentially can leads to the unity in the space $L^2(\dR_+, d\nu(x))$:
\begin{eqnarray}\label{unityOneTwo}
&& B_L=\int_{\Group{ Aff}(\dR)} d\mu_L(q,p) \Ket{q,p}\Bra{q,p} \, ,\\
&& B_R=\int_{\Group{ Aff}(\dR)} d\mu_R(q,p) \Ket{q,p}\Bra{q,p} \, ,\\
&& B_U= \int_{\Group{Aff}(\dR)} d\mu_U(q,p) \, \rho(q,p) \Ket{q,p}\Bra{q,p} \, .
\end{eqnarray}
Let us check which one is invariant under the action $U(q,p)$ of the affine
group:
\begin{equation} \label{BLInvU}
U(q',p') B_U U(q',p')^\dagger =
\int_{-\infty}^{+\infty} dq \int_0^\infty dp \rho(q,p)
\Ket{q/p'+q',p'p}\Bra{q/p'+q',p'p}
\end{equation}
One needs to replace the variables under the integral:
\begin{eqnarray} \label{BLInvU2}
&&\tilde{q}=q/p'+q' \quad \mbox{and} \quad \tilde{p}=p'p \, ,\\
&& q=p'(\tilde{q}-q') \quad \mbox{and} \quad p=\frac{\tilde{p}}{p'}\, .
\end{eqnarray}
Calculating the Jacobian
$\frac{\partial(q,p)}{\partial(\tilde{q},\tilde{p})}=1$ one gets:
\begin{equation} \label{BLInvU3}
d\mu_U(q,p)=\rho(q,p)  d\tilde{q}d\,\tilde{p}=
\rho(p'(\tilde{q}-q'),\frac{\tilde{p}}{p'}) d\tilde{q}d\,\tilde{p}
\end{equation}
This implies, the transformed weight should be equal to the initial one,
$\rho(p'(\tilde{q}-q'),\frac{\tilde{p}}{p'})=\rho(q,p)$ for every
$(q',p')$. The simplest solution is $\rho(q,p)=\mathrm{const}$, so we get
$d{\mu_U}(q,p)=dq\,dp$.

It also implies that the operators $B_L$ and $B_R $ do not commute with the
affine group. The action \eqref{im1} is not compatible neither with the left
invariant, nor with right invariant measures on the affine group.

The {\it irreducibility} of the representation, used to define the coherent
states \eqref{im22}, enables making use of Schur's lemma \cite{BR}, which leads
to the resolution of the unity in $L^2(\dR_+, d\nu(x))$:
\begin{equation}\label{im44}
\int_{\Group{ Aff}(\dR)}  d{\mu_U}(q,p) \Ket{q,p}\Bra{q,p} = A_\phi\;\I \; ,
\end{equation}
where the constant $A_\phi$ can be calculated using any arbitrary, normalized
vector $\Ket{f} \in L^2(\dR_+, d\nu(x))$:
\begin{equation}\label{im3b}
A_\phi = \int_{\Group{ Aff}(\dR)}d{\mu_U}(q,p)\,
\BraKet{f}{q,p}\BraKet{q,p}{f} \, .
\end{equation}
This formula can be calculated directly:
\begin{eqnarray}\label{im3b2}
&& A_\phi = \int_{\Group{ Aff}(\dR)}  d{\mu_U}(q,p) \nonumber \\
&& \int_0^\infty d\nu(x')\int_0^\infty d\nu(x)
(f(x')^\star e^{iqx'}\phi(x'/p))(e^{-iqx}\phi(x/p)^\star f(x)) \nonumber \\
&& = \int_0^\infty \frac{dx'}{x'}\int_0^\infty \frac{dx}{x}
\int_0^\infty dp
\left[\frac{1}{2\pi}\int_{-\infty}^{+\infty} dq e^{iq(x'-x)}\right]
f(x')^\star f(x) \phi(x'/p)\phi(x/p)^\star \nonumber \\
&& = \int_0^\infty \frac{dx}{x^2} |f(x)|^2
\int_0^\infty dp |\phi(x/p)|^2    \nonumber \\
&& = \left(\int_0^\infty \frac{dx}{x} |f(x)|^2\right)
\left(\int_0^\infty \frac{dp}{p^2} |\phi(p)|^2 \right)
=\int_0^\infty \frac{dp}{p^2} |\phi(p)|^2 \,
\end{eqnarray}
if $\BraKet{f}{f}=1$.

\noindent In the derivation of \eqref{im3b2} we have used the equations:
\begin{equation}\label{rem}
  \langle x | x^\prime\rangle =
x \delta (x - x^\prime),~~~\int_0^\infty \frac{dx}{x}\,|x \rangle \langle x |
= \id,
~~~\int_0^\infty \frac{dx}{x}\,\delta (x - x^\prime)f(x) = f(x^\prime) \, .
\end{equation}
%

\section{Orthonormal basis of the carrier space} \label{basis}

The  basis of the Hilbert space $L^2(\dR_+, d\nu(x))$ is known  to be \cite{GM}
\begin{equation}\label{aa1}
e^{(\alpha)}_n (x) =
\sqrt{\frac{n!}{(n + \alpha)!}}\,e^{-x/2} x^{(1 +\alpha)/2}\,L_n^{(\alpha)}(x),
\end{equation}
where $L_n^{(\alpha)}$ is the Laguerre polynomial, $\alpha > -1$, and $(n +
\alpha)! = \Gamma (n + \alpha + 1)$.  One can verify that $\int_0^\infty
e^{(\alpha)}_n (x) e^{(\alpha)}_m (x) d\nu(x)= \delta_{n m}$ so that
$e^{(\alpha)}_n (x)$ is an orthonormal basis.


\end{document}